%% file: paper.tex
\documentclass[sigconf]{acmart}

\AtBeginDocument{%
  \providecommand\BibTeX{{%
    \normalfont B\kern-0.5em{\scshape i\kern-0.25em b}\kern-0.8em\TeX}}}

\setcopyright{none}
\renewcommand\footnotetextcopyrightpermission[1]{}

\usepackage{tabularx}
\usepackage{xcolor}
\usepackage[binary-units]{siunitx}
\usepackage{setspace}
\usepackage{acronym}
\usepackage{rotating}
\usepackage[tight]{subfigure}
\usepackage{units}
\usepackage{booktabs}
\usepackage{multirow}
\usepackage{listings}
\usepackage{pgfplots}
\usepackage{pgf-pie}
\usepackage[flushleft]{threeparttable}
\newacro{VM}{virtual machine}
\newacro{MPU}{memory protection unit}

\begin{document}

\title[Quantum-Resistant Security for Software Updates on Low-power Networked Embedded Devices]{Quantum-Resistant Security for Software Updates\\ on Low-power Networked Embedded Devices}

\author{Gustavo Banegas}
\authornote{These authors contributed equally to this work.}
\email{gustavo@cryptme.in}
\affiliation{%
  \institution{Inria and Laboratoire d’Informatique de l’École
  Polytechnique,\\ Institut Polytechnique de Paris}
  \city{Palaiseau}
  \country{France}
}

\author{Koen Zandberg}
\authornotemark[1]
\email{koen.zandberg@inria.fr}
\affiliation{%
  \institution{Inria}
  \city{Saclay}
  \country{France}
}

\author{Adrian Herrmann}
\email{adrian.herrmann@fu-berlin.de}
\affiliation{%
  \institution{Freie Universität Berlin}
  \city{Berlin}
  \country{Germany}
}

\author{Emmanuel Baccelli}
\email{emmanuel.baccelli@inria.fr}
\affiliation{%
  \institution{Inria}
  \city{Saclay}
  \country{France}
}

\author{Benjamin Smith}
\email{smith@lix.polytechnique.fr}
\affiliation{%
  \institution{Inria and Laboratoire d’Informatique de l’École
  Polytechnique,\\ Institut Polytechnique de Paris}
  \city{Palaiseau}
  \country{France}
}

\renewcommand{\shortauthors}{Banegas and Zandberg and Herrmann and Baccelli and Smith}

\begin{abstract}
\input{0-abstract}

\end{abstract}



\maketitle
\pagestyle{plain}

\input{1-intro}
\input{2-related}
\input{3-case-study}

\input{4-quantum-resistant-sig-landscape}
\input{5-benchmarks}
\input{6-discussion}
\input{7-conclusion}

\begin{acks}
H2020 SPARTA and the RIOT-fp project partly funded this work.
\end{acks}
\

\bibliographystyle{ACM-Reference-Format}
\bibliography{bibliography}

\end{document}

%% file: 0-abstract.tex
As the Internet of Things (IoT) rolls out today to devices
whose lifetime may well exceed a decade,
conservative threat models should consider attackers with access to quantum computing 
power. 
The SUIT standard (specified by the IETF) defines a security 
architecture for IoT software updates, standardizing the metadata and 
the cryptographic tools---namely, digital signatures and hash functions---that guarantee the legitimacy of software updates.
While the performance of SUIT has previously been evaluated in the pre-quantum context,
it has not yet been studied in a post-quantum context.
Taking the open-source implementation of SUIT available in RIOT
as a case study, we overview post-quantum considerations, and
quantum-resistant digital signatures in particular,
focusing on low-power, microcontroller-based IoT devices 
which have stringent resource constraints in terms of memory, CPU, and energy 
consumption.
We benchmark a selection of proposed post-quantum signature schemes
(LMS, Falcon, and Dilithium) 
and compare them with current pre-quantum signature schemes (Ed25519 and ECDSA).
Our benchmarks are carried out on a variety of IoT hardware including ARM Cortex-M, RISC-V, and 
Espressif (ESP32), which form the bulk of modern 32-bit microcontroller architectures. 
We interpret our benchmark results in the context of SUIT,
and estimate the real-world impact
of post-quantum alternatives for a range of typical software update categories.

%% file: 1-intro.tex
\section{Introduction}

Decades of experience with the Internet and networked software has brought about the motto \emph{you can't secure what you can't update}.
Meanwhile, recent technological and societal trends have fuelled the massive deployment of cyberphysical systems;
these systems are increasingly pervasive, and we are increasingly dependent on their functionalities.
A so-called Internet of Things (IoT) emerges, weaving together an extremely wide variety of machines (embedded software and hardware) which are required to cooperate via the network, at large scale.

Unpatched (or worse: unpatchable) devices quickly become liabilities.
Exploits weaponizing compromised IoT devices are demonstrated time and again, sometimes spectacularly as with botnets such as Mirai~\cite{mirai,antonakakis2017mirai}.
The twist is, however, that the cure can become a disease: software updates are themselves an attack vector. Attacks can consist in lacing a legitimate software update with malware, compromising the updated device in effect~\cite{newman2018inside}.
Once IoT devices are deployed, up and running, it thus becomes crucial to have answers ready to the following questions:
\begin{itemize}
\item How/when is software embedded in IoT devices updated?
\item How are software updates secured?
\item What level of security is provided?
\end{itemize}

In this paper we tackle these questions, focusing on low-power IoT devices, and anticipating attackers which may possess quantum-computing power.

\ \\
{\bf Low-power IoT characteristics.}
A prominent and particularly challenging component of IoT deployments consists in integrating low-power, resource-constrained IoT devices into the distributed system.
These devices are typically based on low-cost microcontrollers (e.g., ARM Cortex M, RISC-V, ESP), which interconnect via a low-power radio (e.g., BLE, IEEE 802.15.4, LoRa) or via a wired communication bus.
An estimated 250 billion microcontrollers are in use today around the globe~\cite{250billionMCU}.
Compared to microprocessor-based devices, microcontrollers aim for a different trade-off:
They offer much smaller capacity in computing, networking, memory~\cite{rfc7228},
in order to achieve radically lower energy consumption and a tiny price tag (<1\$ unit price).
To give an idea, it is not uncommon to have a memory budget of 64
\si{\kilo\byte} of RAM and 500 \si{\kilo\byte} of ROM (flash) in total
for the whole embedded system software---including drivers, crypto libraries, OS kernel, network stack and application logic.
Nonetheless, the functionalities and services provided by constrained
microcontroller-based devices are as crucial as those of other, less constrained elements in the cyberphysical system.

\ \\
{\bf Quantum adversaries.}
Robust and commoditized quantum computers still sound futuristic today.
While it would be hazardous to predict the imminence of a breakthrough,
progress in this domain has picked up substantially.
Among others, prominent Big Tech (including Google, IBM, Intel, Microsoft) have already been designing, building and operating small quantum computers over the last years.
Currently, the capacity of existing quantum computers are being incrementally enhanced with more quantum bits, which steadily improves their performance.
As the lifetime of IoT devices rolled out today can largely exceed a decade,
conservative threat models should consider attackers which benefit from quantum-computing power,
on top of traditional computing power.

\ \\
{\bf Post-quantum cryptography.}
Post-quantum cryptosystems
(and in particular, post-quantum signature schemes)
are designed to run on contemporary
hardware, yet resist adversaries equipped with both classical and
quantum computers.
There are many signature schemes that claim post-quantum security,
some old and some new,
but until now none has seen wide deployment.
Recent international research on post-quantum schemes
has revolved around the 
National Institute of Standards and Technology 
(NIST) Post-Quantum Cryptography project~\cite{NIST-PQC},
which aims to distinguish a limited number of candidate schemes
for eventual standardization.
This process is currently in its third round,
and draft standards are expected between 2022 and 2024.

\ \\
{\bf Quantum-resistant security for low-power IoT.}
Let's get back to the motto \emph{you can't secure what you can't update (securely)}.
In our quest for quantum-resistant security, a first order priority
is to guarantee in a future-proof manner that software updates received via the network on low-power IoT devices
are legitimate.
When verifying the legitimacy of a software update,
the crucial cryptographic tool is a digital signature.
Open standards targeting IoT security (such as the IETF~\cite{tschofenig2019cyberphysical})
specify the safe usage of a variety of digital signature schemes to secure software updates on low-power devices,
including one scheme (LMS~\cite{rfc8554}) that offers quantum resistance.

The questions that we aim to answer in this paper are:
\begin{itemize}
    \item \emph{How do post-quantum security costs compare to typical
        pre-quantum security costs?}
    \item \emph{What is the footprint of quantum resistance security,
        relative to typical low-power operating system footprints?}
    \item \emph{What are the potential alternatives for post-quantum
        digital signature schemes to secure IoT software updates?}
    \item \emph{What hash functions should be used in this context?}
\end{itemize}
We will address these questions under the assumption that we want to
achieve, and maintain, a standard level of 128-bit conventional security
(matching current internet security standards)
and NIST Level 1 post-quantum security.

%% file: 2-related.tex
\section{Related Work}

The performance of pre-quantum digital
signature schemes in the context of secure software updates on various
Cortex-M microcontrollers is evaluated in~\cite{zandberg2019suit}.
Various NIST candidate post-quantum schemes
are compared as component algorithms in TLS~1.3 in~\cite{SikeridisKD20},
analyzing performance, security, and key and signature sizes,
as well as the impact of post-quantum authentication on TLS~1.3
handshakes in realistic network conditions,
while \cite{cloudfare_tls}
shows a real life experiment with clients using 
two post-quantum schemes:
an isogeny-based algorithm (SIKE) and a lattice-based algorithm (HRSS).
More recently, another experiment with different schemes was conducted by
Cloudflare~\cite{cloudfare_kem,SchwabeSW20}.

For pure post-quantum cryptographic implementation work targeting
microcontrollers,
\cite{CamposKRS20} evaluates the performance of stateful LMS on
Cortex-M4 microcontrollers,
while \texttt{pqm4}~\cite{pqm4} aims to implement and benchmark 
NIST candidate schemes on Cortex-M4, with M4 assembly subroutines plugged into
some of the \texttt{PQClean} implementations.
(Note that among the NIST candidate signature schemes,
\texttt{PQClean} implements only Dilithium, Falcon, Rainbow, and SPHINCS+;
of these, \texttt{pqm4} implements only Dilithium and Falcon.)
Software verifying SPHINCS, RainbowI, GEMSS,
Dilithium2, and Falcon-512 signatures in Cortex-M3 using less than
8~\si{\kilo\byte} of RAM is presented in~\cite{cryptoeprint:2021:662}.

Many post-quantum digital signature schemes use SHA3 hashing primitives under the hood.
While the performance of SHA3 implementations in hardware (FGPA) have been studied in work such as ~\cite{kaps2011lightweightSHA3fpga,jungk2011area,guo2010fair},
surprisingly few studies focus on the performance of software implementations
of this standard hashing primitive 
on low-power microcontrollers.
Some prior work exists in this domain with particular focus on 8-bit microcontrollers~\cite{balasch2012compactSHA3-8bit,kim2020efficientSHA3-8bit}.
Another prior study has focused on comparing the performance of Keccak variants on 32-bit ARM Cortex-M microcontrollers in ~\cite{adrian_thesis}.

\paragraph{Paper contributions \& outline}
The main contributions of this paper are:
\begin{itemize}
\item We provide an overview of the SUIT specification for secure software updates on low-power IoT devices, using its open-source implementation in a common operating system (RIOT) as case study; We show how
crypto primitives including digital signatures and hash functions are used in compliance with SUIT;
\item We analyze post-quantum considerations for SUIT-compliant hash functions, which we benchmark on low-power 32-bit microcontrollers;
\item We survey quantum-resistant digital signature schemes, and derive
a selection of schemes most applicable for the secure IoT software update
use case;
\item We benchmark signature schemes on heterogeneous low-power IoT hardware
based on popular 32-bit microcontrollers (ARM Cortex-M, RISC-V and ESP32);
\item We compare the performance of quantum-resistant digital signature schemes
(LMS, Dilithium, and Falcon) against that of typical pre-quantum schemes (Ed25519, and secp256);
\item We conclude on the cost of post-quantum security, and outline
perspectives for low-power IoT.
\end{itemize}

%% file: 3-case-study.tex
\section{Case Study: Low-power Software Updates with SUIT}

The IETF standardizes the SUIT specifications (Software Updates for Internet of Things~\cite{suit-manifest, rfc9019}), which define a security architecture, standard metadata and cryptographic schemes able to secure IoT software updates, applicable on microcontroller-based devices.
An open-source implementation of the SUIT workflow is for example available in RIOT~\cite{suit-riot}, a common operating system for low-power IoT devices~\cite{baccelli2018riot} which we use as base for our case study.

\subsection{SUIT Workflow}

\begin{figure*}[t]
  \centering
    \includegraphics[width=1.5\columnwidth]{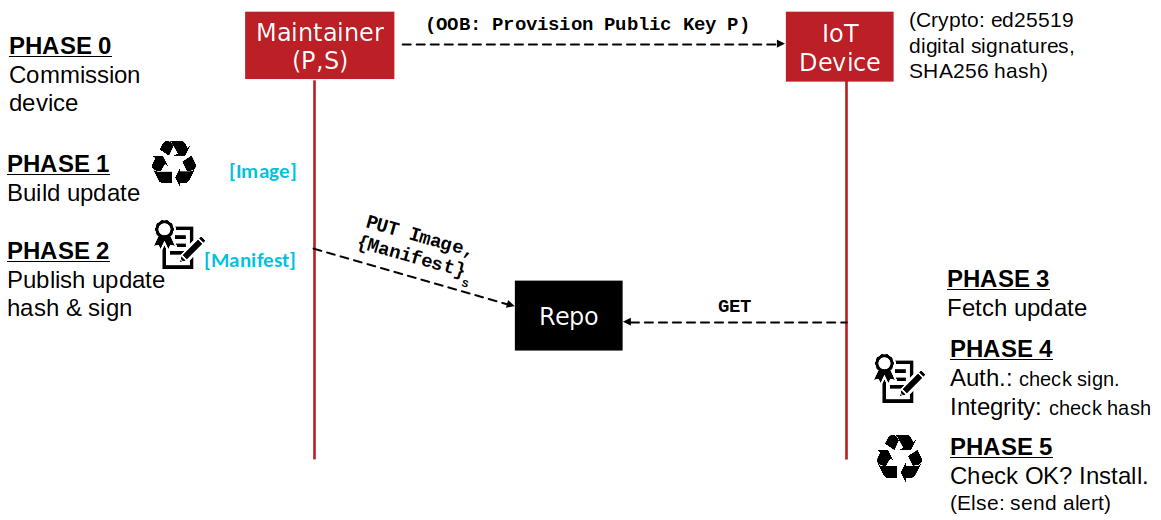}
  \caption{SUIT secure software update workflow.}
  \label{fig:suit-workflow}
\end{figure*}

The SUIT workflow is shown in Fig.~\ref{fig:suit-workflow}. This workflow consists of a preliminary phase (\emph{Phase 0}) whereby the authorized maintainer produces and flashes the IoT devices with commissioning material: the bootloader, the initial image, and authorized crypto material.

Once the IoT device is commissioned, up and running, the maintainer can trigger iterations through cycles of phases 1-5, whereby the authorized maintainer can build a new image (\emph{Phase 1}), hash and sign the corresponding standard metadata (the so-called SUIT manifest, \emph{Phase 2}) and transfer to the device over the network via a repository (e.g. a CoAP resource directory). The IoT device can then fetch the update and the SUIT manifest from the repository (\emph{Phase 3}), proceed to verify the signature and the hash (\emph{Phase 4}), and upon successful verification, the new software is installed and booted (\emph{Phase 5}), otherwise the update is dropped.

Using the metadata and the cryptographic primitives as specified by SUIT, the IoT device can mitigate attacks using software updates as vector, such as:
\begin{itemize}
\item {\bf Tampered/Unauthorized Firmware Update Attacks:} An attacker may try to update the IoT device with a modified and intentionally flawed firmware image. To counter this threat, SUIT specifies the use of digital signatures on a hash of the image binary and the metadata to ensure integrity of both the firmware and its metadata.
\item {\bf Firmware Update Replay Attacks:} An attacker may try to replay a valid, but old (known-to-be-flawed) firmware. This threat is mitigated by using a sequence number used in the metadata, which is increased with every new firmware update.
\item {\bf Firmware Update Mismatch Attacks:} An attacker may try replaying a firmware update that is authentic, but for an incompatible device. To counter this threat, SUIT specifies the inclusion of device-specific conditions, which can be verified before installing a firmware image, thereby preventing the device from using an incompatible firmware image.
\end{itemize}
For a more complete documentation of attacks that are mitigated with SUIT we refer readers to ~\cite{suit-information-model-12}.

\paragraph{SUIT Cryptographic Tools}
As depicted in Fig.~\ref{fig:suit-workflow}, cryptographic tools on which software updates in general and SUIT in particular rely are a digital signature scheme and a hash function.
On the one hand, the digital signature authenticates the software update binary.
On the other hand, to make the signature verification less cumbersome, the signature is not performed on the
software update binary itself, but on a hash of the software update binary.
Thus, this hash function is also a crucial cryptographic primitive in the SUIT workflow.
Fig.~\ref{fig:suit-workflow} depicts this workflow combining SHA-256 hashing and Ed25519 signatures.

\subsection{Hash Functions with SUIT}
The metadata of the update (specified by the SUIT
Manifest~\cite{suit-manifest}) includes a cryptographic hash of the sofware update binary.
To be considered secure, a cryptographic hash function 
$H: \{0,1\}^{*} \to \{0,1\}^{l}$, where $l>0$, 
must have the following properties:
\begin{itemize}
	\item Preimage resistance: Given a hash value $h$, it should be
        infeasible to find any input $m$ such that $h = H(m)$;
	\item Second preimage resistance: Given an input $m_1$, it
        should be infeasible to find a different input $m_2$ such that $H(m_1) = H(m_2)$;
	\item Collision resistance: It should be infeasible to find two
        different inputs $m_1$ and $m_2$ such that $H(m_1) = H(m_2)$.
\end{itemize}

The SUIT standard specification~\cite{suit-manifest}
allows for the use of the following hash functions:
\begin{itemize}
\item SHA-2: either 224-, 256-, 384-, or 512-bit output;
\item SHA-3: either 224-, 256-, 384-, or 512-bit output.
\end{itemize}

\paragraph{Background on SHA-2}
This algorithm is a well-known secure hash function developed in the early 90's
and standardized in 2001 by NIST in~\cite{fips180}. SHA-2 is based on
the Merkle-Damg\r{a}rd construction, and executes several rounds of a
compression function.
SHA-2 presents $4$ different digest sizes (i.e., output lengths):
$224$, $256$, $384$, and $512$ bits.
SHA-2 is widely used in several applications,
including TLS, SSL, PGP, SSH, and many others.
The main reason for this is that it is a stable and secure function:
The best known attacks against SHA-2 break preimage resistance for 52 out of 64 rounds (for
SHA-256), and 57 out of 80 rounds (for SHA-512).
For collision resistance, the only attack is against 46 out of 64 rounds of SHA-256.

\paragraph{Background on SHA-3}
This hash algorithm was standardized in 2015 by
NIST in FIPS 202~\cite{fips202}. SHA-3's basic primitive is called Keccak~\cite{DBLP:conf/eurocrypt/BertoniDPA13}, which
is built using the sponge construction
(differing from SHA-2's Merlke--Damg\r{a}rd construction).
This gives SHA-3 the advantage of resisting known attacks on
Merkle--Damg\r{a}rd hash functions (like SHA-2).
Like SHA-2, SHA-3 presents $4$ different digest sizes: $224$, $256$, $384$, and $512$ bits.
Sponge-based functions can ``absorb'' any amount of data,
and then ``squeeze'' any amount of output,
thus providing an extendable output function (XOF) that can be used for
purposes beyond just hashing.
The Keccak-based XOF specified in FIPS 202 is called SHAKE.

\subsubsection{Post-Quantum Considerations}
\ \\
There are few quantum attacks against SHA-2 and SHA-3 in literature.
Recently,~\cite{BanegasB17} showed that it is possible to parallelize Grover's algorithm
to find preimages of hash functions, and this attack applies to both
Merkle--Damg\r{a}rd hashes (e.g.~SHA-2) and Sponge-based hashes (e.g.~SHA-3).
For collision resistance, the state-of-the-art
in quantum collision search does not drastically reduce the complexity
with respect to classical algorithms, as shown in~\cite{ChaillouxNS17}.
On the other hand, classical attacks for SHA-2 might become a reality,
as shown in~\cite{DobraunigEM15a}.

\subsubsection{Low-Power IoT Considerations}
\ \\
Low-power systems should be able to run hash functions
in the smallest amount of time and using as little power as possible.
Minimal memory (RAM and flash) usage are also desirable.
In this context, since we aim for 128-bit security, the two functions we should consider for SUIT are SHA-256 and SHA3-256.

\begin{figure*}[t]
  \centering
  \begin{minipage}{\columnwidth}
    \centering
    \includegraphics[width=\columnwidth]{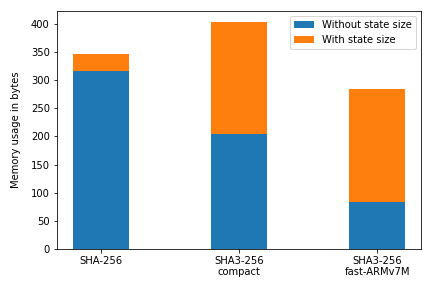}
    \caption{RAM (stack) usage of SHA2 and SHA3 on an ARM Cortex-M4 microcontroller.}
    \label{fig:memory-sha3}
  \end{minipage}%
  \hfill
  \begin{minipage}{\columnwidth}
    \centering
    \includegraphics[width=\columnwidth]{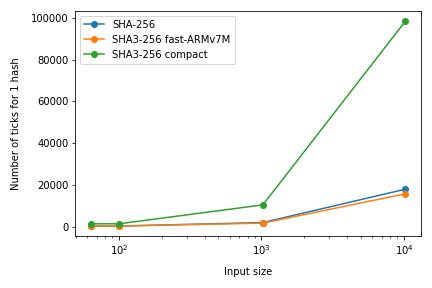}
    \caption{Execution speed of SHA2 and SHA3, versus input size in bytes, on an ARM Cortex-M4 microcontroller.}
    \label{fig:performance-sha3}
  \end{minipage}
\end{figure*}

Table~\ref{tab:flash-sha3} shows the amount of flash memory taken up by
RIOT's default implementation of SHA-256, and compares it with the footprint of two different SHA3-256 implementations:
one optimized for flash memory, and the other optimized for speed on an ARM Cortex-M4 microcontroller (ARMv7M architectures).
Next, Figure~\ref{fig:performance-sha3} shows the execution speed of hash operations using these different implementations on an ARM Cortex M4 microcontroller.
Finally, Figure~\ref{fig:memory-sha3} compares the RAM (stack) memory used by these implementations. 
We observe that RAM usage is roughly equivalent across the different
implementations, but speed and flash can vary widely for the SHA-3 implementations.
Basically, SHA3-256 can offer slightly faster execution compared to SHA-256,
but at the price of a $10\times$ larger flash footprint.
For a flash footprint similar to SHA-256, the comparative speed of SHA3-256 diminishes drastically for larger inputs.
For more detailed analysis, we refer readers to a previous study~\cite{adrian_thesis} comparing different Keccak variants on microcontrollers.

\begin{table}[ht]
  \caption{ROM (flash memory) footprint of SHA2 and SHA3 on an ARM Cortex-M4 microcontroller.}
  \label{tab:flash-sha3}
	\centering
	\bgroup
	\begin{tabular}[ht]{ l | c c }
    \toprule
		\emph{SHA-256} & 1008 bytes \\
		\hline
		\emph{SHA3-256 compact} & 1692 bytes \\
		\emph{SHA3-256 fast-ARMv7M} & 11548 bytes \\
    \bottomrule
	\end{tabular}
	\egroup
\end{table}

\subsubsection{Conclusions on SUIT Hash Functions for Post-Quantum}
\label{sec:hash-conclusions}
\ \\
Based on our analysis, there are no \emph{direct} post-quantum aspects to consider here.
Rather, the decision of which hash function to use should be driven by low-power criteria, and by other \emph{indirect} post-quantum aspects detailed below.

Let's distinguish broad categories for low-power IoT software updates:
\begin{enumerate}
\item Software module update (\textasciitilde5\si{\kilo\byte})
\item Small firmware update without crypto (\textasciitilde50\si{\kilo\byte})
\item Small firmware update with crypto (\textasciitilde50\si{\kilo\byte})
\item Large firmware update (\textasciitilde250\si{\kilo\byte})
\end{enumerate}
In cases (1) and (2), the updated software does not include the hash
function implementation (the cryptographic tools are external, e.g., in a bootloader).
In such cases, the flash memory footprint overhead for the hash function in use is of no concern, and SHA3-256 (optimized for speed) is the best choice.
In cases (2) and (3), however,
the updated software includes the cryptographic tools and the hash function code;
thus, a tradeoff appears.
For small firmware updates, a 10 \si{\kilo\byte} flash overhead represents a significant 25\% bump in what needs to be stored on the device and transmitted over the network.
As updates are infrequent, execution speed may be considered less of a priority, and thus both SHA-256 and flash-optimized SHA3-256 are valid options.
For larger updates, the storage and transfer overhead is negligible, so speed-optimized SHA3-256 is the best option again.

Let us now consider a complementary perspective: We observe that most of the quantum-resistant digital signature schemes use SHA-3 in their constructions. 
In fact, the NIST competition candidates for the upcoming post-quantum
signature standard are required to be SHA-3/SHAKE compatible,
because that is the current US standard.
In this respect, as code footprint on IoT devices is very limited, factorization is typically desirable: There is an opportunity to implement a single hash function (used both for hashing and for signing) in order to use less flash memory.

For these reasons, we SHA3-256 is the primary choice in our case-study.

\subsection{Digital Signatures with SUIT}
The SUIT architecture relies on the software update distributor
(i.e., the authorized maintainer in Figure~\ref{fig:suit-workflow})
issuing a long-term public-private key pair,
and the public key being pre-installed on the IoT device(s) to be updated,
during the commissioning (\emph{Phase 0}) shown in Fig.~\ref{fig:suit-workflow}).
This key pair is used to generate and verify a digital signature on the IoT software update.
Digital signature use in SUIT
is specified in the COSE standard~\cite{rfc8152},
which defines how to sign and encrypt compact (CBOR) binary serialized objects.
COSE standardizes the use of elliptic-curve digital signature schemes
using the following state-of-the-art elliptic curves:
\begin{itemize}
\item Curve25519 (Ed25519), Curve448 (Ed448);
\item NIST P-256, P-384, P-521.
\end{itemize}

\begin{figure}
\begin{tikzpicture}[scale=0.8]
 \pie[polar, explode=0.1, text=legend]
    {14/Crypto,
     32/Kernel,
     38/Network stack,
     16/OTA module
     }
\end{tikzpicture}
  \caption{Flash memory compostion of SUIT-enabled RIOT firmware with typical pre-quantum configuration using Ed25519 signatures and SHA-256 hashing (total footprint 52.5 \si{\kilo\byte}).}
\label{pie:mem:total-system}
  \label{fig:suit-footprint}
\end{figure}
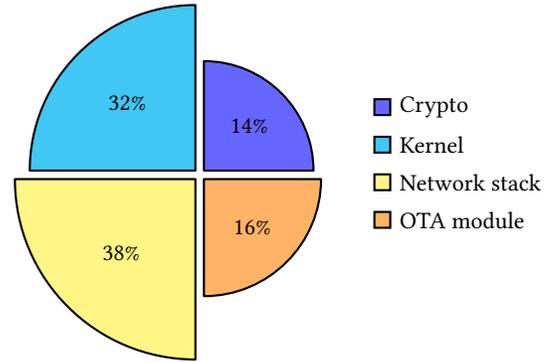

These elliptic-curve schemes are desirable
as they offer very small public (and private) keys
at 32 bytes each and 64-byte signatures.
To give some concrete perspective,
Figure~\ref{fig:suit-footprint} shows the memory footprint of SUIT and related software components when using Ed25519, compared to the whole software embedded on the IoT device.
For this measurement
we used the available open-source RIOT implementation,
which we build for,
and run on a popular low-power IoT board 
based on an ARM Cortex-M4 microcontroller
(the Nordic nRF52840 Development Kit).
The flash memory footprint of this firmware is 52.5 \si{\kilo\byte} and the RAM (stack) usage is
16.3 \si{\kilo\byte}.

In particular, we observe that in this typical pre-quantum
configuration, out of a $\approx50$ \si{\kilo\byte} total flash footprint,
the crypto represents a minor part of the footprint (under 15\%).
Furthermore, the SUIT manifest itself can remain small, because the elliptic-curve signature adds 15\% to the size of the SUIT manifest metadata and less than 0.1\% to the data that must be transferred over the network, counting the manifest and the firmware binary as depicted in \autoref{tab:network-transfer-cost-SUIT-Ed25519}.

\begin{table}[ht]
  \caption{SUIT firmware update network transfer cost, using minimal metadata, Ed25519 signatures, SHA-256 hashing.}
  \label{tab:network-transfer-cost-SUIT-Ed25519}
	\centering
	\bgroup
	\begin{tabular}[ht]{ l | c c}
    \toprule
            & Network Payload \\
		\hline
		\emph{SUIT metadata} & 419 bytes \\
		\emph{SUIT signature} & 64 bytes \\
		\emph{OS firmware} & 52485 bytes \\
    \bottomrule
	\end{tabular}
	\egroup
\end{table}


\subsubsection{Post-Quantum Considerations}
\ \\
Elliptic-curve schemes are advantageous
because they provide high security guarantees even though keys and signatures are very small.
However, the security of elliptic-curve signatures
is guaranteed by the hardness of the elliptic-curve Discrete Logarithm
Problem, which can be solved efficiently on large quantum computers
using Shor's algorithm~\cite{Shor94,HanerJNRS20,BanegasBHL21}.

It is important to note that a breakthrough in quantum computing
at a time $T$ will not affect the security of elliptic-curve signatures
generated before $T$,
but it would certainly destroy the security of any elliptic-curve
signatures generated after $T$.
In our use case, the distributor's key pair has a very long planned
lifetime, possibly equal to that of the devices to be updated; securely
updating the key itself will be impossible, or at least undesirable.
We therefore need to build-in resistance to the quantum threat in
anticipation of such a development.

\subsubsection{Low-Power IoT Considerations}
\ \\
The range of post-quantum signature schemes
considered as potential replacements for elliptic-curve signatures
is wide and diverse, and the idiosyncrasies that distinguish the
various schemes are exaggerated by the constraints of low-power IoT devices.
However, all of these schemes have public key and signature sizes that are
one or two orders of magnitude larger than the elliptic-curve
equivalents.
Post-quantum signatures are therefore far from drop-in replacements;
they represent a significant research challenge
for microcontroller and IoT implementations.

Nevertheless, the IETF recently initiated the standardization of alternative signature schemes with COSE/SUIT which aim to offer quantum-resistant security levels, such as LMS~\cite{rfc8554}.
In the next sections, we survey alternative quantum-resistant schemes,
and give an experimental comparison of their performance against that of state-of-the-art pre-quantum digital signature schemes as used in SUIT.

%% file: 4-quantum-resistant-sig-landscape.tex
\section{Post-Quantum Digital Signatures}

The signature schemes that we consider
target at least NIST Level~1 for \emph{post-quantum security}.
This is the basic security level proposed by NIST as
part of its Post-Quantum Cryptography (PQC) Standardization
Project~\cite{NIST-PQC}.
Level 1 security includes both 128 bits of classical security,
and an equivalent level of security with respect
to some model of quantum computation.
That is, an adversary should require on the order of \(2^{128}\)
operations to gain any non-negligible advantage when attacking the
scheme, even if this adversary benefits from quantum computing power.
For example, this is the amount of required work for an adversary
to have any chance of forging a signature on a new message,
under a given public key.
This 128-bit security level is now standard in mainstream internet applications
requiring long-term security.

\subsection{Post-quantum signature paradigms}

Post-quantum signatures,
like other post-quantum protocols,
form natural families according to the sources of underlying hard
problems that guarantee their security:

\paragraph{Hash-based signatures}
Hash-based signatures are among the oldest digital signature schemes.
Their security is based on the difficulty of inverting cryptographic
hash functions. The security assumptions have been well studied, which
gives an academic maturity to the problem. 
Important contemporary examples include XMSS~\cite{rfc8391},
LMS~\cite{rfc8554},
and the NIST Round~3 alternatate candidate {SPHINCS+}~\cite{SPHINCS}.
Hash-based signatures tend to offer very fast verification,
though this comes at the cost of very large signatures.

\paragraph{Lattice-based signatures}
``Lattice-based'' schemes
are based on hard problems in Euclidean lattices,
and related problems like Learning With Errors (LWE).
Contemporary examples of lattice-based signatures
include the NIST Round~3 finalists
Dilithium~\cite{Dilithium},
based on the module-LWE problem,
and Falcon~\cite{Falcon},
based on the NTRU problem.
These schemes offer fast signing and verification, at the cost of relatively large signatures.

\paragraph{Multivariate signatures}
The security of ``multivariate'' schemes
is based on the difficulty of solving certain low-degree polynomial
systems in many variables.
Important contemporary examples of multivariate signatures
include the NIST Round~3 finalist Rainbow~\cite{Rainbow},
and alternate candidate GeMSS~\cite{GeMSS}.
Multivariate signatures like Rainbow and GeMSS are interesting because they offer
extremely small signature sizes, though this comes at the cost of
extremely large public keys.
However, recent analysis as in~\cite{Beullens20a} has brought their
security levels into question.

\paragraph{Isogeny-based signatures}
Isogeny-based cryptosystems
are based on the difficulty of computing unknown isogenies between elliptic curves.
They inherit small key sizes from conventional elliptic-curve
cryptography (ECC),
which makes them interesting for microcontroller applications,
but they also inherit (and increase) ECC's burden of heavy
algebraic calculations.
SIKE~\cite{SIKE}
is a NIST Round~3 alternate candidate for key establishment,
but no isogeny-based signatures were submitted to the NIST PQC
project---for the simple reason that
no reasonable algorithms existed before the project deadline.
Recent proposals such as SQISign~\cite{SQISign}
offer small parameters that are attractive for microcontroller
applications, albeit at the cost of very slow runtimes.
However, these signature schemes have not yet been subjected to extensive security
analysis, and implementation work is still at an early stage.

\paragraph{Code-based signatures}
Code-based cryptosystems are based on the difficulty of
hard problems from the theory of error-correcting codes.
The McEliece key exchange scheme~\cite{McEliece1978}
is among the oldest of all public-key cryptosystems.
Code-based signatures, on the other hand,
are much less well-established.
While some recent proposals such as Wave~\cite{Wave,DebrisAlazardS19}
have interesting potential,
their security analysis and implementation work
lags even further behind that of isogeny-based signatures.

\paragraph{Zero-knowledge-based signatures}
There is also a category of post-quantum signatures
using Zero-Knowledge (ZK) techniques,
combining algorithms from symmetric cryptography
with a technique known as Multi-Party Computation In The Head.
The NIST Round~3 candidate Picnic~\cite{Picnic}
is an important example:
It offers extremely small key sizes,
but at the cost of very large signatures.

\subsection{Selection of candidates}
\label{sec:PQ-selection}

When choosing candidates for evaluation in our use case,
we must consider not only their key and signature sizes
and their runtime performance,
but also their maturity with respect to security analysis.
While the relatively compact parameters of some isogeny- and code-based signature schemes
may make them interesting for future work targeting microcontrollers,
at present these schemes are far from theoretical maturity,
let alone real-world deployability.
The true security level of the NIST multivariate and ZK-based
candidates is a subject of current debate,
though their extremely large keys and/or signatures
would likely eliminate them from consideration
for our applications in any case.

The NIST PQC project has dominated research in post-quantum cryptography in
recent years.
Its candidate cryptosystems are a natural first port of call
for credible post-quantum signature algorithms,
since they have had the benefit of concerted analysis from the
cryptographic community---especially the Round 3 proposals,
which are candidates for standardization in the coming years.
However, these are not the only algorithms that we should consider.
For example, among hash-based signature schemes,
we might compare the older LMS scheme (which is not a NIST candidate)
with the newer SPHINCS+ scheme (which is a NIST Round~3 alternate).
LMS has smaller computational requirements,
but the signer must maintain some state between signatures;
SPHINCS+ is a heavier scheme, but it is stateless.
Statelessness is an advantage for general applications.
In our use case, however, statefulness is natural
(it corresponds naturally to the version number on the software update),
and easier to maintain---so the lighter LMS is a more natural choice.

\paragraph{Post-quantum choices}
For the reasons above,
we chose to focus our efforts on three post-quantum signature
algorithms: LMS, Dilithium, and Falcon,
representing the hash-based and lattice-based categories.
LMS has 60-byte public keys and 4756-byte signatures.
Dilithium III, targeting NIST security level 2,
has 1312-byte public keys and 2420-byte signatures.
Falcon-512, targeting NIST security level 1,
has 897-byte public keys and 666-byte signatures.

\paragraph{Pre-quantum choices}
To make a meaningful comparison with pre-quantum algorithms,
we selected two elliptic-curve schemes:
the state-of-the-art Ed25519~\cite{Ed25519paper,rfc8032} scheme,
and the historic standard ECDSA based on the secp256 curve~\cite{FIPS186-4}.
These schemes offer particularly small public keys and signatures---just
32 and 64 bytes, respectively---with an acceptable runtime and memory footprint
for applications targeting microcontroller platforms.

%% file: 5-benchmarks.tex
\section{Benchmarks}\label{sec:benchmark}

\subsection{Hardware Testbed Setup}
We carried out our measurements on popular, commercial, off-the-shelf IoT hardware.
Our chosen platforms are representative of the landscape of modern 32-bit microcontroller architectures,
including ARM Cortex-M, Espressif ESP32 and RISC-V:
\begin{itemize}
    \item a Nordic nRF52840 Development Kit, which provides
        a typical microcontroller (ARM Cortex-M4)
        with $256$ \si{\kilo\byte} RAM,
        $1$ \si{\mega\byte} flash,
        and a $2.4$ \si{\giga\hertz} radio transceiver
        compatible with both IEEE 802.15.4 and Bluetooth Low-Energy.
    \item a WROOM-32 board (ESP32 module with the ESP32-D0WDQ6 chip on board),
        which provides two low-power
        Xtensa\textsuperscript{\tiny\textregistered} 32-bit LX6
        microprocessors with integrated Wi-Fi and Bluetooth,
        $520$ \si{\kilo\byte} RAM,
        $448$ \si{\kilo\byte} ROM
        and $16$ \si{\kilo\byte} RTC SRAM.
    \item a Sipeed Longan Nano GD32VF103CBT6 Development Board,
        which provides a RISC-V 32-bit core
        with $32$ \si{\kilo\byte} RAM
        and $128$ \si{\kilo\byte} flash.
\end{itemize}
IoT-Lab~\cite{iot-lab2015-ieee-iot-wf} provides some of the hardware for reproducibility on open access testbeds.

\subsection{Software Setup}

We used RIOT~\cite{riot:home} as a base for our benchmarks.

\paragraph{Pre-quantum implementations.}
We used three different libraries, all currently supported in RIOT.
\begin{description}
    \item[Ed25519:]
        For Ed25519, we used two libraries: \textbf{C25519} (provided
        in~\cite{c25519}) and \textbf{Monocypher}~\cite{monocypher:home}.
        The C25519 implementation contains constant-time finite-field arithmetic
        based on public-domain implementations of Bernstein’s Curve25519
        key exchange~\cite{bernstein2006curve25519}.
        The Monocypher library also provides an implementation of
        Bernstein’s Curve25519 and the Ed25519 signature scheme.
        One difference between the Monocypher and C25519 implementations
        is that Monocypher uses precomputed tables to speed up the computation
        of elliptic curve points.
        The precomputations are used in
        the ``window method'' for scalar multiplication.
        While this is known to speed up computations, it also has its drawbacks,
        as pointed out in~\cite{longa_ecc}.
    \item[ECDSA:]
        For ECDSA, we used the Intel's \textbf{Tinycrypt} library~\cite{tinycrypt},
        which is designed for embedded devices.
        The main goal of this library is to provide cryptographic standards for constrained devices.
        ECDSA differs from Ed25519 both in some
        specific details of the signature algorithm
        and in using the NIST standard p256 curve instead of Curve25519.
\end{description}

\paragraph{Post-quantum implementations.}
We re-used publicly available code after making some small modifications to fit the hardware requirements.
\begin{description}
    \item[LMS:]
        For LMS, we used the Cisco implementation~\cite{lms_git},
        removing calls to \texttt{malloc}
        (which is not supported by RIOT).
        For our benchmark, we used the smallest parameters proposed
        in~\cite[Section 5]{rfc8554} i.e., we use SHA-2 with $256$-bit output as the hash function,
        with tree height $5$, and $32$ bytes associated with each node.
    \item[Dilithium:]
        Our Dilithium implementation
        was based on the PQClean implementation~\cite{pqclean}.
        As the Dilithium specification~\cite[Sec.~3.1]{Dilithium} states,
        the first step in \emph{signing} and \emph{verifying}
        is to expand a random seed given in the public key into a large matrix.
        For our benchmarks, we prepared two versions of Dilithium:
        \begin{itemize}
            \item
                \textbf{Dynamic Dilithium}
                is the basic PQClean implementation.
            \item
                \textbf{Static Dilithium}
                is a modification of the PQClean implementation
                where the matrix is precomputed
                and stored in the flash memory.
        \end{itemize}
        Precomputing and storing the matrix
        makes signing and verification both faster,
        though it also (by definition) requires more flash
        and reduces flexibility,
        since signatures can only be verified against the flashed key.
    \item[Falcon:]
        We used the the Falcon implementation provided by
        PQClean~\cite{pqclean},
        without any significant structural modifications.
\end{description}

\subsection{Measurements}

\begin{table}[t]
    \caption{Private key, public key, and signature size
    for each of the evaluated signature schemes.}
\label{tab:sizes}
\resizebox{\columnwidth}{!}{
\begin{tabular}{rr|rrr}
\toprule
 & Algorithm & Private Key (\si{\byte}) & Public Key (\si{\byte})& Signature (\si{\byte}) \\
\midrule
\multirow{2}{*}{Pre-quantum}
  & Ed25519         & 32    & 32   & 64    \\
  & ECDSA p256   & 32    & 32   & 64    \\
\midrule
\multirow{3}{*}{Post-quantum}
 & Falcon     & 1281    & 897   & 666    \\
 & Dilithium  & 2528    & 1312  & 2420    \\
 & LMS (RFC8554)        & 64      & 60    & 4756    \\
\bottomrule
\end{tabular}
}
\end{table}

\begin{table*}
  \begin{minipage}{\textwidth}

    \begin{center}
      \caption{Benchmark of pre/post-quantum signature schemes using an ARM Cortex-M microcontroller (nRF52840 Dev. Kit).}
      \label{tab:benchNRF}
      \resizebox{\textwidth}{!}{
      \begin{tabular}{rrr|rrr|rrr}
          \toprule
              & & & \multicolumn{3}{c}{Sign} & \multicolumn{3}{c}{Verify}
          \\
          \cmidrule(r){4-6} \cmidrule(r){7-9}
              & \multirow{2}{*}{Algorithm} & Flash
              & \multicolumn{2}{c}{Time} & Stack
              & \multicolumn{2}{c}{Time} & Stack
          \\
              & & (\si{\byte}) & (\si{\milli\second}) & (KiloTicks) & (\si{\byte}) & (\si{\milli\second}) & (KiloTicks) & (\si{\byte})
          \\
          \midrule
          \multirow{3}{*}{Pre-quantum}
              & Ed25519 (C25519)      & 5106  & 845  & 54111  & 1180  & 1953 & 125012 & 1300  \\
              & Ed25519 (Monocypher)  & 13852 & 17   & 1136   & 1420  & 40   & 2599   & 1936  \\
              & ECDSA p256 (Tinycrypt)& 6498  & 294  & 18871  & 1084  & 313  & 20037  & 1024  \\
          \midrule
          \multirow{4}{*}{Post-quantum}
              & Falcon                & 57613 & 1172 & 75020  & 42240 & 15   & 1004   & 4744  \\
              & Dilithium (Dynamic)   & 11664 & 465  & 29788  & 51762 & 53   & 3407   & 36058 \\
              & Dilithium (Static)    & 26672 & 135  & 8655   & 35240 & 23   & 1510   & 19504 \\
              & LMS (RFC8554)         & 12864 & 9224 & 590354 & 13212 & 123  & 7908   & 1580  \\
          \bottomrule
          \vspace{0.2cm}
      \end{tabular}}
  \end{center}
  \end{minipage}
    \begin{minipage}{\textwidth}
      \begin{center}
      \caption{Benchmark of pre/post-quantum signature schemes using an Espressif ESP32 microcontroller (WROOM-32 board).}
      \label{tab:benchESP}
      \resizebox{\textwidth}{!}{
      \begin{tabular}{rrr|rrr|rrr}
          \toprule
              & & & \multicolumn{3}{c}{Sign} & \multicolumn{3}{c}{Verify}
          \\
          \cmidrule(r){4-6} \cmidrule(r){7-9}
              & \multirow{2}{*}{Algorithm} & Flash
              & \multicolumn{2}{c}{Time} & Stack
              & \multicolumn{2}{c}{Time} & Stack
          \\
              & & (\si{\byte}) & (\si{\milli\second}) & (KiloTicks) & (\si{\byte}) & (\si{\milli\second}) & (KiloTicks) & (\si{\byte})
          \\
          \midrule
          \multirow{3}{*}{Pre-quantum}
              & Ed25519 (C25519)      & 5608  & 921  & 73690  & 1312  & 2165 & 173205 & 1440  \\
              & Ed25519 (Monocypher)  & 17238 & 21   & 1709   & 1536  & 60   & 4864   & 2160  \\
              & ECDSA p256 (Tinycrypt)& 6869  & 333  & 26696  & 1296  & 374  & 29948  & 1216  \\
          \midrule
          \multirow{4}{*}{Post-quantum}
              & Falcon                & 60358 & 1172 & 93824  & 42504 & 16   & 1322   & 4920  \\
              & Dilithium (Dynamic)   & 12397 & 87   & 7036   & 51954 & 43   & 3508   & 36242 \\
              & Dilithium (Static)    & 27197 & 121  & 9694   & 35412 & 21   & 1706   & 19620 \\
              & LMS (RFC8554)         & 15177 & 7583 & 606674 & 13488 & 101  & 8141   & 1808  \\
          \bottomrule
          \vspace{0.2cm}
      \end{tabular}
      }
    \end{center}
    \end{minipage}
    \hfill
    \begin{minipage}{\textwidth}
      \begin{center}
      \caption{Benchmark of pre/post-quantum signature schemes using a RISC-V microcontroller (Sipeed Longan Nano board).}
      \label{tab:benchNano}
      \begin{threeparttable}
        \resizebox{\textwidth}{!}{
      \begin{tabular}{rrr|rrr|rrr}
          \toprule
              & & & \multicolumn{3}{c}{Sign} & \multicolumn{3}{c}{Verify}
          \\
          \cmidrule(r){4-6} \cmidrule(r){7-9}
              & \multirow{2}{*}{Algorithm} & Flash
              & \multicolumn{2}{c}{Time} & Stack
              & \multicolumn{2}{c}{Time} & Stack
          \\
              & & (\si{\byte}) & (\si{\milli\second}) & (KiloTicks) & (\si{\byte}) & (\si{\milli\second}) & (KiloTicks) & (\si{\byte})
          \\
          \midrule
          \multirow{3}{*}{Pre-quantum}
              & Ed25519 (C25519)      & 6024  & 956  & 68883  & 1312  & 2242 & 161475 & 1440  \\
              & Ed25519 (Monocypher)  & 17328 & 16   & 1194   & 1376  & 41   & 3013   & 1920  \\
              & ECDSA p256 (Tinycrypt)& 7452  & 270  & 19489  & 1224  & 308  & 22192  & 1112  \\
          \midrule
          \multirow{3}{*}{Post-quantum}
              & Falcon                & 11122\tnote{1} & ---  & ---    & ---   & 13   & 975    & 4756  \\
              & Dilithium (Dynamic)   & ---   & ---  & ---    & ---   & ---  & ---    & ---   \\
              & Dilithium (Static)    & 25148\tnote{2} & ---  & ---    & ---   & 17   & 1237   & 19572 \\
              & LMS (RFC8554)         & 15889 & 9105 & 655614 & 13352 & 122  & 8808   & 1736  \\
          \bottomrule
      \end{tabular}
      }
      \begin{tablenotes}
      \item[1] Flash only contains the verification algorithms.
      \item[2] Flash contains the verification algorithms and hard-coded keys.
    \end{tablenotes}
  \end{threeparttable}
\end{center}
    \end{minipage}
\end{table*}

Table~\ref{tab:sizes} gives the sizes (in bytes) of the private key, public key, and signature
for each of the schemes that we measured.

Tables~\ref{tab:benchNRF}, ~\ref{tab:benchESP}, and ~\ref{tab:benchNano} present
our benchmarking results on three different architectures Cortex-M, ESP32 and RISC-V.

For ease of comparison, all of the tables follow the same format.
The upper and lower halves of each table
describe the pre- and post-quantum algorithms, respectively.
For each scheme and implementation tested,
we give the total flash memory (ROM) used by the library;
then, for the signing and verification operations
we list the running time in milliseconds as well as in (thousands of)
``ticks'', which we computed from the hardware clock and time spent.
We also report the amount of stack memory required to successfully run
the operation.

In Table~\ref{tab:benchNRF} we see that Monocypher's Ed25519
is the fastest for signing
among all the candidates running on the Nordic board;
but Falcon is the fastest for signature verification,
followed by Static Dilithium.

Table~\ref{tab:benchESP} shows that
Monocypher's Ed25519 is also the fastest to sign on the WROOM-32 board,
while Falcon and Static Dilithium offer the fastest signature verification.

Table~\ref{tab:benchNano} gives results on a RISC-V processor.
Since the RISC-V board has only 32 \si{\kilo\byte} RAM,
the Falcon and Dilithium signing algorithms could not be run.
For signature verification, we can see that
the post-quantum schemes are substantially faster
than the pre-quantum schemes.

%% file: 6-discussion.tex
\section{The impact of quantum resistance in SUIT/COSE}

With our empirical results from~\S\ref{sec:benchmark},
we can now address the three outstanding questions
posed in the introduction of this paper
(having already addressed hash functions in~\S\ref{sec:hash-conclusions}).

\subsection{The cost of post-quantum security}

\emph{How do post-quantum security costs compare to typical
pre-quantum security costs?}
A toe-to-toe comparison between pre-quantum and
post-quantum algorithms must consider
public key and signature sizes,
running time,
and memory requirements.

Table~\ref{tab:sizes} shows
that post-quantum algorithms always have larger
public key and signature sizes,
generally by well over an order of magnitude.
Compared with standard elliptic-curve signature schemes,
Falcon's public keys are
$28\times$ larger
and its signatures are $10.4\times$ larger;
Dilithium's public keys are
$41\times$ larger than elliptic-curve keys,
and its signatures are $38\times$ larger.
LMS avoids this spectacular growth in public key sizes,
with keys only $1.875\times$ larger than elliptic-curve public keys;
but its signatures are a massive $74.3\times$ larger than elliptic-curve signatures.

Looking at running time,
as we saw in \S\ref{sec:benchmark},
post-quantum signatures have their
advantages and disadvantages.
Signature verification is considerably faster
across all the IoT devices that we tested.
Signing is generally slower.
A comparison of the signing algorithms in Table~\ref{tab:benchNRF}
shows that the fastest post-quantum algorithm runs in $135$ ms,
which is $7.94\times$ slower than Ed25519 (Monocypher).
But the tables are turned when we
compare signature verification algorithms:
The fastest pre-quantum algorithm runs in $40$ ms,
which is $2.65\times$ slower than post-quantum Falcon.
Efficient verification is a required and valuable feature
(in all scenarios),
but in this setting, it comes at the price of
an increase in stack and flash memory.

Looking at memory requirements,
we see that post-quantum flash requirements can grow to over
$11\times$ the smallest pre-quantum flash.
Similarly, post-quantum algorithms impose a considerable
increase in stack memory.

Moving to post-quantum signatures therefore entails
an increase in memory resources (stack and flash)
and bandwidth (for keys and signatures).
However, the verification algorithms are faster than standard pre-quantum
algorithms, which means a reduction both in latency and in energy consumption.
Ultimately, when choosing between
these signature schemes in practice, one must
consider the target IoT device, update frequency,
and bandwidth usage.

\subsection{The cost of post-quantum SUIT/COSE}

\emph{What is the footprint of quantum-resistant security, relative
to typical low-power operating system footprints?}
To add quantum resistance to SUIT/COSE
following the workflow presented in Figure~\ref{fig:suit-workflow},
we change the cryptographic algorithms
from Ed25519 and SHA256
to Falcon, LMS, or Dilithium, and SHA3-256.

\paragraph{Impact on the SUIT Manifest}
In practical terms,
what changes in the size of the manifest in Phase~2,
is an increase according to the signature size.

For example:
Suppose we build a firmware update for RIOT, for the nRF52840dk board (based on a Cortex-M4 microcontroller).
The protocol requires publishing the manifest containing the signed hash
of the image: That is, we need to publish both the image and the SUIT metadata containing the signature.
In \S\ref{tab:network-transfer-cost-SUIT-Ed25519} we had measured that the SUIT manifest with pre-quantum Ed25519 (or ECDSA) is of a total size of $419 + 64 = 483$ \si{\byte}.
Moving to post-quantum signatures, this total becomes
$419 + 666 = 1085$ \si{\byte} for Falcon,
$419 + 2420 = 2839$ \si{\byte} for Dilithium,
and $419 + 4756 = 5175$ \si{\byte} for LMS.
Comparing these post-quantum sizes with
the pre-quantum baseline,
we see that post-quantum signatures are
$\approx2.24\times$ bigger with Falcon,
$\approx5.87\times$ bigger with Dilithium,
and $\approx9.84\times$ bigger with LMS.

\paragraph{Impact on the SUIT Software Update Performance}

In order to grasp the whole picture, one must also consider the crucial aspect of network transfer costs.
Based on our results, we evaluate the cost of the entire SUIT software update process in \autoref{tab:suit-cost-post-quantum}.
We distinguish two main cases, as we did in~\S\ref{sec:hash-conclusions}.
In the first, the updated software does \emph{not} include the cryptographic
libraries binary (i.e., these tools are external, e.g., in a
bootloader): this corresponds to the column \emph{Transfer} in
\autoref{tab:suit-cost-post-quantum}.
In the second, the updated software includes the cryptographic libraries
binary; the column \emph{Transfer w. crypto} corresponds to this case.

As we can observe in \autoref{tab:suit-cost-post-quantum}, the impact of switching to quantum-resistance security level on the SUIT update process varies widely in terms of network transfer costs, ranging from negligible increase (~1\%) to major impact ($3\times$ more), depending on the software update use case.

\begin{table*}[ht]
  \caption{Relative cost increase for SUIT with quantum resistance (on ARM Cortex M-4).}
  \label{tab:suit-cost-post-quantum}
	\centering
	\bgroup
	\begin{tabular}[ht]{ l | c c c c}
    \toprule
        SUIT & Flash & Stack & Transfer & Transfer w. crypto \\
		\hline
		\emph{base w. Ed25519 / SHA256}  & 52.4\si{\kilo\byte} & 16.3\si{\kilo\byte} & 47\si{\kilo\byte} & 53\si{\kilo\byte}\\
		\hline
		\emph{with Falcon / SHA3-256}    & +120\% & +18\% & +1.1\% & +120\% \\
		\emph{with LMS / SHA3-256}       & +34\% & +1.2\% & +9\% & +43\%  \\
		\emph{with Dilithium / SHA3-256} & +30\% & +210\% & +4.3\% & +34\%  \\
    \bottomrule
	\end{tabular}
	\egroup
\end{table*}

\subsection{Post-quantum signatures for IoT}
\emph{What are the potential alternatives for post-quantum digital
signature schemes to secure IoT software updates?}
There are many possible deployments of IoT,
and several possible scenarios for IoT software updates.
The authorized maintainer is responsible for updating the firmware,
and it is safe to assume that the maintainer has a PC, or equivalent powerful hardware.
Hence, the computational burden of signing is not the main concern here.
On the other hand,
a constrained device will be responsible for signature verification
in Phases~3, 4, and~5,
as Figure~\ref{fig:suit-workflow} shows.
\emph{Our primary focus in the cost/benefit analysis is therefore on signature verification, rather than signing}.

Let us revisit the four prototypical software update categories
from \S\ref{sec:hash-conclusions},
and consider the choice of postquantum signatures for each.

In Cases (1) and (2),
the package will contain the software update and the signature.
Hence, speed and signature size are more important than flash size.
In these cases, \textbf{Falcon} has an advantage over the alternatives we looked at (LMS and Dilithium).

Case (3) is more complicated, with flash playing a much more crucial role.
Since we must transfer the update with crypto over a low-power network,
the package size has a higher impact on energy costs.
As a point of reference, it takes 30s to 1mn to transfer 50\si{\kilo\byte} on a low-power IEEE802.15.4 radio link, depending on the radio link quality, and the network load (assuming non-extreme cases).
This is to compare with plus/minus 2s of computation speed difference for signature verification among candidate cryptographic tools, for instance.
In this case, as shown in \autoref{tab:suit-cost-post-quantum}, \textbf{LMS} presents the best tradeoff
between flash size, network transfer costs, verification time, and stack size, amongst the candidates we studied.

In Case (4), the large network transfer costs will overwhelm the other costs,
reducing the comparative advantages of one post-quantum signature over
another.

From a purely cryptographic point of view, given the
maturity of hash function cryptanalysis,
LMS remains the safest choice.
As we briefly explained in~\S\ref{sec:PQ-selection},
hash-based problems are quite mature, and have
received extensive cryptanalysis from
the cryptographic community.
In comparison, the security of structured lattice-based schemes like Falcon
is less well-understood.

Nevertheless, compared to pre-quantum state of the art, LMS imposes a significant increase in signature size and running time for
signing and verifying, which impacts greatly the performance of SUIT;
thus, despite its relative lack of maturity,
the performance characteristics of Falcon make it extremely tempting for
applications with smaller updates.

%% file: 7-conclusion.tex
\section{Conclusion}

In this paper, we have studied experimentally the use of quantum-resistant cryptography applied to securing software updates on low-power IoT devices.
Taking an open-source implementation of the IETF standard SUIT as concrete case study, 
we offer a direct comparison
of pre-quantum cryptographic schemes (signatures and hashing) against post-quantum cryptographic schemes
in the same environment and on the same hardware platforms.
We evaluate the cost of upgrading the security level
from pre-quantum 128-bit security to 
NIST Level 1 post-quantum security.
To properly analyze the impact of quantum-resistant cryptography schemes in IoT
updates, we overviewed and selected candidate schemes, and we compared their performance using three low-power IoT
platforms (ARM Cortex-M, RISC-V, and ESP32) representative of the current landscape of 32-bit microcontrollers.
We show that quantum resistance is indeed achievable today on these platforms, and we derive recommendations based on our performance analysis.
We also characterize how post-quantum digital signatures take a significant toll on the memory footprint and/or on the network transfer costs incurred in the IoT software update process, compared to pre-quantum schemes.

For future work, the priority remains to stabilize the current
versions of post-quantum signatures, and then to push their implementations to common low-power embedded software platforms such as RIOT.
Meanwhile, NIST is still determining which candidate schemes
might form a new post-quantum signature standard;
should new candidates be included in a new call,
a new analysis will become necessary.

